# Temporal patterns of synchrony in a pyramidal-interneuron gamma (PING) network

Quynh-Anh Nguyen[1], Leonid L Rubchinsky[1,2]

[1]Department of Mathematical Sciences, Indiana University Purdue University Indianapolis, Indianapolis, IN, USA
[2]Stark Neurosciences Research Institute, Indiana University School of Medicine, Indianapolis, IN, USA

**Abstract**

Synchronization in neural system plays an important role in many brain functions. Synchronization in the gamma frequency band (30Hz-100Hz) is involved in a variety of cognitive phenomena; abnormalities of the gamma synchronization are found in schizophrenia and autism spectrum disorder. Frequently, the strength of synchronization is not very high and is intermittent even on short time scales (a few cycles of oscillations). That is, the network exhibits intervals of synchronization followed by intervals of desynchronization. Neural circuits dynamics may show different distributions of desynchronization durations even if the synchronization strength is fixed. In this study, we use a conductance-based neural network exhibiting pyramidal-interneuron (PING) gamma rhythm to study the temporal patterning of synchronized neural oscillations. We found that changes in the synaptic strength (as well as changes in the membrane kinetics) can alter the temporal patterning of synchrony. Moreover, we found that the changes in the temporal pattern of synchrony may be independent of the changes in the average synchrony strength. Even though the temporal patterning may vary, there is a tendency for dynamics with short (although potentially numerous) desynchronizations, similar to what was observed in experimental studies of neural activity synchronization in the brain. Recent studies suggested that the short desynchronizations dynamics may facilitate the formation and the break-up of transient neural assemblies. Thus, the results of this study suggest that changes of synaptic strength may alter the temporal patterning of the gamma synchronization as to make the neural networks more efficient in the formation of neural assemblies and the facilitation of cognitive phenomena.

**Synchronization of neural oscillations is a common neural phenomenon believed to be relevant to a large range of neural functions and dysfunctions. Neural synchrony at rest is rarely perfect and fluctuates in time. Few long desynchronizations and many short desynchronizations may lead to different functional consequences even if the average synchrony strength is not changed. This study explores the potential network mechanisms of different temporal patterns of neural synchrony in the model of synchronized neural gamma oscillations, which are related to cognitive function of the brain. The study shows how gamma rhythm can be partially synchronized with specific temporal patterning and how this temporal patterning of gamma synchronization is regulated by connectivity strength and other factors. Furthermore, the study shows how temporal patterning of neural synchronization can be varied independently of the synchrony strength. Understanding the mechanisms of temporal patterning of neural synchrony may help to understand its relation to neural function.**

# 1. INTRODUCTION

Synchronization in neural networks is a widespread phenomenon that is important for a variety of brain functions and dysfunctions. Through synchrony, collective behavior in neural networks can be established; thus, synchrony may play an important role in memory, cognition, perception (e.g., Buzsáki and Draguhn, 2004). Abnormal synchrony is found to be associated with different brain disorders such as Parkinson's disease (Hammond, et al., 2007; Oswal et al., 2013; Rubchinsky et al, 2012), schizophrenia (Uhlhaas and Singer, 2010; Pittman-Poletta et al., 2015; Spellman and Gordon, 2015), and autism (Sun et al., 2012; Malaia et al., 2020). Synchronization in the gamma frequency band is a focus of many studies as it is believed to be responsible for the facilitation of interneuronal communication for cognition (Fries, 2015).

Synchronization in the brain networks is not a perfect synchronization, at least not at the rest state. While this may be affected by many factors, when a network shows a moderate synchrony strength, it goes in and out of the synchronized state. Networks with similar synchrony strength can have completely different synchrony pattern. One can have many brief desynchronization events or a few long desynchronization events even if the synchrony strength is the same. Given the importance of synchrony in the brain for behavior, the temporal patterning of synchrony on short time scales should be important.

Techniques to detect and analyze the temporal patterning of synchronous dynamics were recently developed (Park et al., 2010; Ahn et. al., 2011), using the first-return maps for the phases of oscillations. These techniques were applied to experimental data (e.g., Park et al., 2010; Ahn et al., 2013, 2014; Ratnadurai-Giridharan et al., 2016; Malaia et al., 2020; Dos Santos Lima et al., 2020); it was found that the patterning of neural synchrony (even if the overall synchrony strength is not changed) may be correlated with behavior (Ahn et al., 2014, 2018; Malaia et al., 2020). One of the interesting observations of all these studies was that the temporal patterning of synchronization was very specific: oscillations go out of synchrony predominantly for very short amount of time (although they may do so rarely or frequently resulting in high or low overall synchrony).

In the present study, we use these analysis techniques to investigate the temporal patterning of synchronization in the gamma frequency band. We consider a model of two connected circuits exhibiting pyramidal-interneuron gamma (PING) rhythm. The properties of gamma rhythm in these circuits rely upon synaptic time scales and synaptic strength of excitatory and inhibitory connections (e.g., Ementrout and Kopell, 1998; Buzsáki and Wang, 2012; Borgers, 2017). We hypothesize that inhibitory and excitatory synaptic connections do not only change gamma oscillations and their synchrony level but also alter the temporal pattern of synchrony. While the network we use may be viewed as a somewhat simplistic representation of gamma rhythm in the brain, our objective is to see if and how synaptic and cellular properties may potentially affect the temporal structure of synchrony as a proof of principle. We found that the temporal patterning of synchrony can be changed by the synaptic and cellular changes and can even be altered independently of the overall synchrony strength. We further conclude with the discussion of the modeling results in the context of available experimental analysis of the temporal patterning of neural synchronization.

## 2. METHODS

Our network consists of two synaptically connected circuits, each of which generates gamma-band activity in isolation. Each circuit includes two excitatory neurons and two inhibitory neurons and is adapted from (Borgers, 2017). Figure 1A,B illustrates the schematic of the network.

### 2.1. Model neurons and synapses

Each model neuron is represented by a single compartment conductance-based model (see, e.g., Izhikevich, 2007; Ermentrout and Terman, 2010). Transmembrane voltage is given by:

$$C_m \frac{dV}{dt} = -I_{Na} - I_K - I_L - I_{syn} + I_{app} \tag{1}$$

with the membrane currents described below. $I_{Na} = g_{Na} m^3 h (V - v_{Na})$ is the transient sodium current. The activation is considered to be instantaneous and $m$ is taken as $m = \alpha_m(V) / (\alpha_m(V) + \beta_m(V))$. The inactivation function $h$ obeys first-order kinetics:

$$\frac{dh}{dt} = \alpha_h(V)(1-h) - \beta_h(V)h \tag{2}$$

$I_K = g_K n^4 (V - v_K)$ is the persistent potassium current, and the activation function $n$ obeys first-order kinetics:

$$\frac{dn}{dt} = \alpha_n(V)(1-n) - \beta_n(V)h \tag{3}$$

Here, $\alpha_*$ and $\beta_*$ are probabilities of opening and closing of the corresponding channel, respectively. Finally, $I_L = g_L(V - v_L)$ is the leak current, and $I_{app}$ is a constant applied current.

Excitatory neurons and inhibitory neurons have different set of parameters and $\alpha_*$ and $\beta_*$ functions. Excitatory neurons follow reduced Traub and Miles model (Traub and Miles, 1991) with $C_m = 1\ \mu F/cm^2, v_{Na} = 50\ mV, v_K = -100 mV, v_L = -67 mV, g_{Na} = 100\ mS/cm^2, g_K = 80\ mS/cm^2, and\ g_L = 0.1\ mS/cm^2$. The $\alpha_*(V)$ and $\beta_*(V)$ functions are given below:

$$\alpha_m(V) = \frac{0.32(V+54)}{1 - \exp\left(-\frac{V+54}{4}\right)} \qquad \beta_m(V) = \frac{0.28(V+27)}{\exp\left(\frac{V+27}{5}\right) - 1} \tag{4}$$

$$\alpha_h(V) = 0.128 \exp\left(-\frac{V+50}{18}\right) \qquad \beta_h(V) = \frac{4}{1+\exp\left(-\frac{V+27}{5}\right)} \qquad (5)$$

$$\alpha_n(V) = \frac{0.032(V+52)}{1-\exp\left(-\frac{V+52}{5}\right)} \qquad \beta_n(V) = 0.5\exp\left(-\frac{V+57}{40}\right) \qquad (6)$$

Inhibitory neurons follow Wang-Buzsáki model (Wang and Buzsáki, 1996) with $C_m = 1\,\mu F/cm^2, v_{Na} = 55\,mV, v_K = -90\,mV, v_L = -65\,mV, g_{Na} = 35\,mS/cm^2, g_K = 9\,mS/cm^2$, and $g_L = 0.1\,mS/cm^2$. The $\alpha_*(V)$ and $\beta_*(V)$ functions are given below:

$$\alpha_m(V) = \frac{0.1(V+35)}{1-\exp\left(-\frac{V+35}{10}\right)} \qquad \beta_m(V) = 4\exp\left(-\frac{V+60}{18}\right) \qquad (7)$$

$$\alpha_h(V) = 0.35\exp\left(-\frac{V+58}{20}\right) \qquad \beta_h(V) = \frac{5}{1+\exp\left(-\frac{V+28}{10}\right)} \qquad (8)$$

$$\alpha_n(V) = \frac{0.05(V+34)}{1-\exp\left(-\frac{V+34}{10}\right)} \qquad \beta_n(V) = 0.625\exp\left(-\frac{V+44}{80}\right) \qquad (9)$$

The synaptic current is given as $I_{syn} = g_{syn}\,s(t)\,(V_{post} - v_{syn})$. Here, $V_{post}$ is the potential of the postsynaptic cell, $v_{syn}$ is the reverse synaptic potential, and $g_{syn}$ is the synaptic strength. Synaptic gating variable $s(t)$ follows first order kinetics equation (Izhikevich, 2007; Ermentrout and Terman, 2010):

$$\frac{ds}{dt} = H(V_{pre})\frac{1-s}{\tau_r} - \frac{s}{\tau_d} \qquad (10)$$

where $H(V_{pre}) = (1 + tanh(V_{pre}/4))/2$ is a sigmoidal function of the presynaptic neuron potential $V_{pre}$, and $\tau_r, \tau_d$ are synaptic rise and synaptic decay time constants, respectively. AMPA-receptor-mediated excitatory synapse has $\tau_r = 0.1\,ms$, $\tau_d = 3\,ms$, $v_{syn} = 0\,mV$. GABA-receptor-mediated inhibitory synapse has $\tau_r = 0.3\,ms$, $\tau_d = 9\,ms$, $v_{syn} = -80\,mV$. These parameters are taken from (Borgers et al., 2012; Borgers, 2017).

## 2.2. Network connectivity

The model is comprised of two circuits with two excitatory neurons (E neurons) and two inhibitory neurons (I neurons) in each circuit. Synaptic connection strength within circuit is denoted as $g_{syn} = g_*$, and $g_{syn} = c_*$ is connection between circuit, see Figure 1. Synaptic connection strength is measured in $mS/cm^2$; for the brevity we will not use the units of measurements when we refer to synaptic strength. It has been pointed out that EE connection should not play a significant role in gamma oscillation (Borgers, 2017; Ermentrout and Kopell, 1998); thus, we set $g_{EE} = c_{EE} = 0$. All other connections between neurons are included in the model, and there are no recurrent connections.

Within each circuit, there are inhibitory synapses between I neurons as well as from I to E neurons, and excitatory synapses from E to I neurons. In the full network (two connected circuits), each E neuron receives inhibitory inputs from two I neurons of the same circuit ($g_{IE}$) and two I neurons of the other circuit ($c_{IE}$). Each I cell receives excitatory inputs from two E neurons of the same circuit ($g_{EI}$) and two E neurons of the other circuit ($c_{EI}$), inhibitory inputs from the other I neuron of the same circuit ($g_{II}$) and two I neurons of the other circuit ($c_{II}$). The schematic of the model is summarized in Fig. 1A,B. The default values for synaptic connections between neurons are $g_{IE} = 0.7$, $g_{EI} = 0.1$, $g_{II} = 0.3$, and $c_{IE} = c_{EI} = c_{II} = 0.02$.

Neurons from each circuit have slightly different values of constant $I_{app}$, so that circuits have slightly different frequencies. In the slower circuit, E neurons have $I_{app} = 4.5 \, mA/cm^2$ and $4 \, mA/cm^2$, while I neurons have $I_{app} = 0.1 \, mA/cm^2$ and $0.09 \, mA/cm^2$. In the faster circuit, E neurons have $I_{app} = 5 \, mA/cm^2$ and $4.5 \, mA/cm^2$, while I neurons have $I_{app} = 0.08 \, mA/cm^2$ and $0.07 \, mA/cm^2$. The slower circuit has 44.4 Hz average firing rate and the faster circuit fires has 46.8 Hz average firing rate.

We vary the strength of different synapses as described below. Some current kinetics parameters are also varied in some numerical simulations as described below. Otherwise, parameters are kept at their default values. The system is solved using adaptive Runge-Kutta (4,5) method (MATLAB ode45 solver) for $25 \, s$.

### 2.3. Time-series analysis

To analyze the dynamics of synchronization between two circuits, we look at the relationship between the phases of oscillators on short time scales (one cycle of oscillations), employing the time-series analysis approach used in earlier experimental studies (Park et al., 2010; Ahn and Rubchinsky, 2013; Ahn et al., 2014) and computational studies (Ahn and Rubchinsky, 2017) of neural synchrony. We consider the time-series of total synaptic current into a neuron. In each circuit, we choose the excitatory neuron that has higher firing rate. First, we use Hilbert transform to compute the phase of each time-series, denoted as $\varphi_1(t)$ and $\varphi_2(t)$. Then, the average synchronization index (e.g., Pikovsky et al., 2004; Hurtado et al., 2004) is computed as:

$$\gamma = \left\| \frac{1}{N} \sum_{j=1}^{N} e^{i(\varphi_1(t_j) - \varphi_2(t_j))} \right\| \tag{11}$$

where $\varphi_1(t_j)$ and $\varphi_2(t_j)$ are the phases of a neuron from circuit 1 (slower circuit) and a neuron from circuit 2 (faster circuit), respectively, at time $t_j$, N is the number of timepoints, and $\|.\|$ denotes the magnitude of a complex number. The average synchronization index $\gamma$ varies between 0 (no synchrony) and 1 (full synchrony). For intermediate values of $\gamma$, the system is partially synchronized.

We then proceed to characterize the temporal pattern of synchronization. The idea of this approach is first to find the presence of a synchronized state (which requires analysis of long time-series, as synchronization is not an instantaneous phenomenon) and then to track the oscillations on each cycle by checking if they are close to the synchronized state or not. The method for it has been described earlier in (Park et al., 2010; Ahn and Rubchinsky, 2013, 2017; Ahn et al., 2014). Briefly, a discrete set of phase difference values $\{\phi_i\}$ is obtained by recording the value of $\varphi_2(t_j)$ whenever $\varphi_1(t_j)$ goes from negative to positive values. For partially synchronized dynamics, these values will cluster around some mean. Note that the mean is not necessarily zero and thus this analysis detects not only zero lag synchronization. If $\{\phi_i\}$ is more than $\pi/2$ away from its mean value, the signals are considered to be in the desynchronized state during the cycle $i$; otherwise, the signals are considered to be in the synchronized state. The number of consecutive cycles in which the signals are desynchronized is called the duration of desynchronization.

The distribution of desynchronization durations provides a statistical description of the temporal patterning of synchronized dynamics. Examples of desynchronization durations distributions are shown in panels E, F, and G of Figures 2-10, where the horizontal axis in histograms measures the duration of desynchronizations in the number of cycles of oscillations. The number of desynchronization durations in each histogram usually varies around 300-600. Following earlier studies of these temporal patterns, we will use mode and a desynchronization ratio, as well as the average desynchronization duration. The mode tells us the most common desynchronization duration, and $f_{mode}$ measures how frequent the modal value is. The value of $f_{mode}$ describes how much the system favors the most typical synchronizations; values closer to 1 show that most of the desynchronizations last as long as their mode. Desynchronization ratio is defined as the ratio of the relative frequency of desynchronized episodes lasting for 1 cycle of oscillations to the relative frequency of desynchronized episodes longer than 4 cycles (similar to how it was used in experimental studies by Ahn et al., 2014, 2018; Malaia et al., 2020). Thus, larger value of desynchronization ratio points to larger number of short desynchronizations.

## 3. RESULTS

In each network, we choose an excitatory neuron that has a higher firing rate and study the synchronized dynamics between them (although, to reflect the network dynamics, we report the average firing frequency in the network; both faster and slower neurons' frequencies are close to each other). With weak to moderate connections between circuits, the network exhibits partially synchronized activity (see Fig. 1C for time-series for an example of voltage and raster plots of spiking in all neurons). This section presents the results of numerical simulations of how synaptic excitation and inhibition and kinetics of ionic channels affect the temporal patterning of this partially synchronized neural oscillations.

### 3.1. Synaptic effects on the temporal pattern of synchronization

*The effect of excitation.*

We consider variation of both within-circuit excitatory connections $g_{EI}$ and cross-circuit excitatory connections $c_{EI}$. We vary $g_{EI}$ from 0.08 to 0.21. In this range, the average firing frequency of the network stays in 44-48 Hz range (Fig. 2A). The synchronization index is in 0.3-0.36 range (Fig. 2B). There is a pronounced change in the mode of the distribution of the desynchronization durations; it changes from 3 to 1 as synaptic strength $g_{EI}$ increases (Fig. 2C). Mode 1 is observed in a relatively large range of $g_{EI}$ (and is typical for various experimental observations). Average desynchronization duration is generally decreasing from about 4 cycles to about 2 cycles (Fig. 2D). Desynchronization ratio (see Methods) shows substantial variation. It increases from about 1 to about 3 with one isolated case of almost 12 (driven by very small value of the relative frequency of long desynchronizations); then it fluctuates in between 2 and 4 (Fig. 2D). We would like to note that a very high value of desynchronization ratio may not necessarily carry a high precision, nevertheless its points to the fact that long desynchronizations are really rare.

Panels E, F, and G show examples of distributions of desynchronization durations for three different values of $g_{EI}$. Example E shows desynchronization distribution with mode 3; the average desynchronization duration is relatively large, and the desynchronization ratio is relatively small. Examples F and G show desynchronization durations distribution with mode 1. While F and G have the same mode, example F has a much higher value of desynchronization ratio because it has a much smaller number of long desynchronizations (lasting 5 and more cycles). The average values of desynchronization durations in both cases are roughly similar. Nevertheless, these examples show a general trend of desynchronizations becoming shorter. These examples corroborate the weak trend described in the previous paragraph: as we increase excitatory synapse strength $g_{EI}$, the desynchronizations become progressively shorter.

Note that mode and average (mean) may have substantially different values. This is probably not surprising (especially given that experimental studies report the mode equal to one, so the average must be larger). It is not clear which particular characteristic of the distribution is of a most biological importance, so mode, desynchronization ratio (as reported in experiments cited in the Introduction) and the average value are presented to better illustrate the dynamics.

Next, we examine the effect of cross-circuit excitatory to inhibitory connection $c_{EI}$ by varying its value from 0 to 0.04. The average firing rate mildly increases from 45 to 47 Hz (Fig. 3A) while the synchronization index stays within 0.3-0.35 range (Fig. 3B). The distribution of desynchronization duration shows substantial changes with different values of $c_{EI}$, see Figure 3C. For smaller values of $c_{EI}$, the mode of the distribution is mostly 1. However, for larger values of $c_{EI}$, the desynchronization durations distribution has predominantly mode 3 (although mode 1 and 2 are also present).

Figure 3D shows how the desynchronization ratio and the average desynchronization duration are changing with $c_{EI}$. They vary around 2 and 3 respectively in the $c_{EI}$ range that produces mostly distributions with mode 1. However, the distribution shows more prominent changes as can be seen in panels E, F, and G: smaller values of $c_{EI}$ (example E) yield a sharper and more prominent mode 1 (more similar to those observed in experiments) than larger values of $c_{EI}$ (example F).

Example G exhibits distribution with the mode 3. Note that even though desynchronization ratio is high for the distribution in G, the mode is not 1. Thus, as we mentioned above, the desynchronization ratio alone is not sufficient to distinguish between short and long desynchronization dynamics.

*The effect of inhibition.*

There are two kinds of inhibitory connections in the model: inhibitory to excitatory neurons and inhibitory to inhibitory neurons connections. We examine the impact of both kinds of inhibitory connections on the temporal patterning of synchronization, looking at within-circuit connections ($g_{IE}$ and $g_{II}$) as well as cross-circuit connections ($c_{IE}$ and $c_{II}$).

Local inhibitory to excitatory connection $g_{IE}$ is varied from 0.6 to 1.36. Within this range, the average frequency decreases significantly from 49 Hz to 36 Hz (Fig. 4A). This is expected; larger $g_{IE}$ results in more inhibition input for excitatory cells and thus reduces the firing rate. The synchronization index varies from 0.3 to about 0.43 (Fig. 4B). The mode of desynchronization durations distribution is mostly 1 (Fig. 4C). There is only one isolated case in which desynchronization duration has higher mode. Furthermore, as $g_{IE}$ increases, the desynch ratio (although variable) shows a generally increasing trend and almost triples from 2 to 6, see Fig. 4D. The average desynchronization duration mildly varies between 2 and 3, see Fig. 4D. Examples of desynchronization durations distribution are shown in panels E, F, and G. Example E shows histogram for the case of small $g_{IE}$ value. While the mode is 1, the likelihood of a desynchronization duration lasting 1 cycle is similar to the likelihood of longer durations. Thus, the desynchronization ratio is low. Examples F and G illustrate histograms for the cases of larger $g_{IE}$ values, and the mode (which is 1) is more prominent in both cases. While desynchronization distributions in F and G have similar average durations, desynchronization ratio in G is noticeably bigger. Thus, in general, larger values of $g_{IE}$ tend to promote shorter desynchronization.

Similarly, we examine the effect of cross-circuit inhibitory to excitatory connection $c_{IE}$. When $c_{IE}$ goes from 0 to 0.08, the average firing rate decreases from 47 Hz to 41 Hz (Fig. 5A), and the synchronization index shows a mild increase from 0.31 to 0.37 (Fig. 5B). The mode of desynchronization durations is mostly 1, and there are a couple isolated cases of mode 2 (Fig. 5C). Further, the average mode increases, and desynchronization ratio decreases (Fig. 5D). Thus, the desynchronizations have a weak tendency of becoming longer as the cross-circuit inhibitory coupling goes up. Examples in panels F and G both show distribution with mode 1; however, mode 1 in example F is more prominent than mode 1 in example G.

Within-circuit inhibitory to inhibitory connection $g_{II}$ is varied from 0.14 to 1. The average frequency varies minimally in between 44 Hz and 46.5 Hz, see Fig. 6A. The synchronization index first increases and then decreases, but stays in between 0.29 and 0.33, see Fig. 6B. Desynchronization durations distribution shows some substantial variations, see Fig. 6C. The higher mode cases all occur for higher values of $g_{II}$. Desynchronization ratio decreases sharply from 12 to 2, and then it stays around 1 to 2; the average desynchronization duration increases slowly from 2 to 4, see Fig. 6D. Example in panel E portrays distribution with a prominent mode 1 and a small probability of any duration lasting 5 cycles or more; thus, the desynchronization ratio is large. Examples F and G shows distributions for larger values $g_{II}$. In this region, the distribution can either have a higher mode (mode 3 in example F) or a lower mode 1 (example G). Either way, the distribution is flatter and thus, larger values of $g_{II}$ tend to produce longer desynchronization durations.

Between-circuit inhibitory to inhibitory connection $c_{II}$ is varied from 0 to 0.11. The average frequency decreases from 46 to 43 Hz, see Fig. 7A. The synchronization index varies in the range of 0.32 and 0.26, see Fig. 7B. The distribution of desynchronization durations has mode 1 in all cases, see Fig. 7C. The average values of desynchronization durations appear to have a very weak decreasing trend while the desynchronization ratio has a weak increasing trend, see Fig. 7D. Panels E, F, and G portray examples of desynchronization durations distributions for small and large values of $c_{II}$. In the first example E, probability of desynch duration lasting 1 cycle is close to the probability of longer cycles. In panels F and G, the likelihood of desynchronization duration lasting 1 cycle is larger than that of longer durations. Also, mode 1 in panel G is more prominent than mode 1 in panel F. Overall, larger $c_{II}$ shows a weak tendency for shorter desynchronizations.

*Changes in the desynchronization durations can be independent of the frequency and of the average synchronization strength.*

In previous sections, we see that changes in the desynchronization durations are often accompanied by the changes in the frequency of oscillations and the average synchronization strength. Here, we consider situations, where synchrony and frequency are not changing while desynchronization durations are. We would like to note that since the time-series analysis method used here measures desynchronization durations in relative units (cycles of oscillations), it is interesting to see what happens when frequency is fixed (otherwise, changes in the desynchronization durations as measured in cycles may not necessarily translate into the same changes in desynchronization duration measured in the absolute time units). To keep frequency and synchronization index relatively constant, yet to alter the temporal pattern of synchrony, we co-vary multiple synaptic strengths. Specifically, we increase within-circuit connections ($g_{IE}$ and $g_{EI}$) and decrease between-circuit connections ($c_{IE}$ and $c_{EI}$) at the same time, parametrizing all of them as a linear function of parameter $k$:

$$g_{EI} = 0.0012k + 0.096 \quad g_{IE} = 0.0041k + 0.8205 \quad for \quad k = 1,2,\dots 31 \quad (12)$$

$$c_{EI} = -0.0004k + 0.038 \quad c_{IE} = -0.0008k + 0.076 \quad for \quad k = 1,2,\dots 31 \quad (13)$$

Within this range of parameters, the average frequency is relatively constant around 40 Hz (Fig. 8A); synchronization index is near 0.37 and does not show substantial variation either (Fig. 8B). When $k$ is small (weak local connections and strong cross-circuit connections), desynchronization durations distribution has mode equal to 3 (Fig. 8C). When $k$ is larger (strong local connections and weak cross-circuit connections), desynchronization durations distribution has mode equal to 1 (Fig. 8C). The average desynchronization duration goes down while the desynchronization ratio shows a prominent increase, pointing to desynchronizations getting shorter (Fig. 8D). This trend is further illustrated in panels E, F, and G with examples of desynchronization durations distributions. Example E has mode 3. Examples F and G have mode 1, and mode 1 is more prominent in example G than in example F. We see that the temporal patterning of synchronized activity may show very substantial changes while average synchrony and frequency are relatively constant. In other words, average synchronization and temporal patterning can be independent of each other.

## 3.2. The effect of membrane current kinetics on the temporal pattern of synchronization

An earlier modeling study in a simple network of two mutually excitatory coupled simplified Hodgkin-Huxley - like model neurons (Ahn and Rubchinsky, 2017) showed that temporal patterning of synchronization is sensitive to the parameters defining the time scale of the delayed rectifier potassium current (responsible for a relaxational character of spiking oscillations in that model). Hence, we want to explore if the model used here (a more complicated network structure with excitation and inhibition and more adequate models of individual neurons) shows a similar dependence of desynchronization durations on the membrane current kinetics.

Similar to (Ahn and Rubchinsky, 2017), we will look at the effect of the peak value of the voltage-dependent activation time-constant and the width of voltage-dependence of the activation time-constant of potassium current (both parameters effectively make this current faster or slower to activate, either directly or indirectly) on the temporal patterning of synchronization. The activation time constant of potassium channel is given by

$$\tau_n(V) = \frac{1}{\alpha_n(V) + \beta_n(V)} \tag{14}$$

where $\alpha_n(V)$ and $\beta_n(V)$ are the opening and closing function of potassium channel (see Equations (6) and (9) in Methods). We parametrize $\alpha_n(V)$ and $\beta_n(V)$ as follows.

$$\alpha_n(V) = \frac{\varepsilon\,\alpha_1(V+52)}{1 - \exp\left(-\frac{V+52}{\delta\,\alpha_2}\right)} \qquad \beta_n(V) = \varepsilon\,\beta_1 \exp\left(-\frac{V+57}{\delta\,\beta_2}\right) \tag{15}$$

Here, $\alpha_1, \alpha_2, \beta_1, \beta_2$ are default values of the opening and closing functions (see Method). Provided that everything else is fixed, varying parameter $\varepsilon$ leads to change of the amplitude of the activation time-constant $\tau_n(V)$. Larger values of $\varepsilon$ lead to faster activation of potassium delayed rectifier current and thus less spiky (closer to sinusoidal waveform) profile of voltage. On the other hand, changing parameter $\delta$ results in change of the width of the activation time-constant $\tau_n(V)$. Larger values of $\delta$ also lead to sinusoidal waveform of voltage trace.

The value of $\varepsilon$ is varied from 0.4 to 3 for all neurons in both circuits. Naturally, this change of the time scale affects the firing rate of the neurons (it moves up from 42 Hz to 51 Hz), see Fig. 9A. The synchronization index stays in the range of 0.3 to 0.35, see Fig. 9B. The shape of the spike is affected too). When $\varepsilon$ is small, the mode of desynch distribution is 1. For larger values of $\varepsilon$, the mode of the desynchronization durations distribution may be 2 or 3; if the mode is 1 in this region, it is not as prominent as mode 1 from small values of $\varepsilon$ (Fig. 9C). Average desynchronization duration is increasing while the desynchronization ratio is decreasing as $\varepsilon$ becomes bigger (Fig. 9D). Examples of the desynchronization durations distributions (shown in panels E, F, and G) illustrate this trend. Thus, slow activation of potassium current (as in the original model, i.e.

physiologically realistic and slower) leads to dynamics with shorter desynchronizations. This is consistent with prior study (Ahn and Rubchinsky, 2017).

The value of $\delta$ is varied between 0.4 and 1.25 for all neurons in both circuits. The average frequency is barely affected by this change, staying in between 45-46 Hz (Fig. 10A). The synchronization index decreases from 0.47 to 0.33 (Fig. 10B). While the mode of desynchronization durations distribution is always 1, the prominence of the mode consistently decreases (Fig. 10C). Fitting the same trend, the average desynchronization duration decreases, and the desynchronization ratio increases (Fig. 10D). Three examples of desynchronization durations distributions shown in panels E, F, and G further illustrate this. Mode 1 is more prominent when $\delta$ is small (example E) as compared to when $\delta$ is large (examples F and G). The distribution in panel G has a smaller mode 1 than the distribution in panel F. Thus, smaller $\delta$ and resulting slower activation of the delayed-rectifier potassium current produces shorter desynchronizations. This is also in agreement with prior study (Ahn and Rubchinsky, 2017).

## 4. DISCUSSION

### 4.1. Summary and significance of findings: connectivity strength affects temporal patterning of network synchronization

We studied the properties of synchronized dynamics of a neural network consisting of two circuits exhibiting PING gamma rhythm. For moderate connectivity strength, gamma oscillations are only partially synchronized, and thus the intervals of highly synchronous activity are interspersed with intervals of low synchrony activity. We found that the temporal patterning of this synchronized activity depends on the strength of connections in the network. Thus, changing synaptic strength affects the distribution of desynchronization events duration (affects the relative duration of the intervals during which the activity is not synchronized).

More specifically, local connections and cross-circuit connections have opposite effects on the temporal pattern of synchronization/desynchronization. Stronger local connections between inhibitory and excitatory neurons (both E to I and I to E synapses within circuits) and weaker cross-circuit connections between inhibitory and excitatory neurons (both E to I and I to E synapses between circuits) promote dynamics with predominantly short desynchronizations. These trends are observed for both independent and simultaneous variation of these synapses. The situation is inversed for connections between inhibitory interneurons. Weaker local connections between inhibitory cells (I to I connections within circuits) and stronger cross-circuit connections between inhibitory cells (I to I connections between circuits) promote dynamics with predominantly short desynchronizations.

We also considered the effect of membrane current kinetics on the temporal patterning of synchronization. While the current kinetics may be harder to change in experiment than synaptic strength, prior modeling studies with minimal neural circuits indicated that it may affect fine temporal structure of neural synchronization (Ahn and Rubchinsky, 2017). We found that different ways of slowing the kinetics of delayed rectifier potassium current (which make a neuron a more relaxational oscillator and lead to more spiky profile of neural voltage) facilitate short desynchronizations.

Furthermore, we showed that temporal pattern of synchronization can vary independently of the average synchronization strength. At the same time, the firing frequency can be kept almost constant so that changes of the desynchronization durations are apparent not only in relative time units (cycles of oscillations), but also in absolute time units (milliseconds). Thus, even though average synchronization strength and temporal pattern of synchrony can co-vary together, they can also vary independently of each other and are independent characteristics of synchronized phenomena in neural networks.

**4.2. Computational results in the context of experimental studies: prevalence of short desynchronization dynamics and the role of synaptic coupling**

Even though the distribution of durations of desynchronization events was found to depend on the properties of neurons and synapses, a review of all the numerical results of this study suggests that partially synchronized dynamics in the PING gamma network has predominantly short desynchronizations. These short desynchronized intervals may be numerous so that the average synchrony level may be low. This is not necessarily true for a generic oscillatory network; the same level of synchrony may be reached with many short desynchronizations and a few long desynchronizations (Ahn et al., 2011, Ahn and Rubchinsky 2017).

However, in the realistic (in the dynamics of neurons and synapses) network studied here, there is a tendency for short desynchronization dynamics. Importantly, the application of the same time-series analysis techniques as used here to various recordings of the electric activity of the brain indicates that it is essentially always dominated by short desynchronizations regardless of the brain area, type of recording, disease status, and brain rhythm. That was observed in the beta band spiking units, LFP, and EEG in Parkinson's disease and its animal model in the basal ganglia and motor cortex (Park et al., 2010; Ratnadurai-Giridharan et al., 2016; Ahn et al., 2018; Dos Santos Lima et al., 2020), alpha and beta band in EEG in healthy subjects (Ahn et al., 2013), theta band in prefrontal cortex and hippocampus in normal and drug-sensitized rodents (Ahn et al., 2014), theta, beta, and low frequency gamma in EEG in subjects with and without autism spectrum disorder (Malaia et al., 2020). The results of the present study provide an additional support for the hypothesis that short desynchronizations dynamics may be common in the synchronization of the oscillations of the neural activity of the brain.

The experimental studies discussed above also found that the changes in the temporal pattern of neural synchrony may be related to behavioral changes, and this may be true even if the average synchronization strength is not changed (Ahn et al., 2014, 2018; Malaia et al., 2020). Even though most of these studies did not consider gamma rhythm, what we found out in the present study of neural network with PING gamma may fit with this general framework of importance of the fine temporal structure of synchronized dynamics. We found that the synchrony pattern may vary if the synaptic strength is varied (and it may vary independently of the synchronization strength). PING gamma is known to depend on the synaptic strength of different types of synapses involved (e.g. Buszaki and Wang, 2012; Salkoff et. al., 2015; Borgers, 2017). A disease with marked abnormalities in the gamma rhythm synchronization, schizophrenia, is known to have alterations in the synaptic strength, in particular, abnormalities in inhibition and in excitatory/inhibitory balance (e.g., Lewis et al., 2005; Vierling-Claassen, et al., 2008; Lisman, 2012; Murray et al., 2014; Grent-'t-Jong et al., 2018). Our study shows these abnormalities may not only affect the average synchronization strength, but may also affect the temporal patterning of synchrony, which, in turn, may affect how neural circuits process information (Ahn and Rubchinsky, 2013, 2017).

### 4.3. Some limitations of the study

There are several limitations of our modeling analysis that we would like to mention here. The model network is quite simple. Of course, no model is perfect, but it is important to remember that our model is a relatively small network while biologically realistic gamma probably requires large networks (Borgers et al., 2012). Also, our model does not consider conduction delays, which may both promote and weaken synchronization (e.g., Woodman and Canavier, 2011). The signals we use here as a proxy for local field potentials are (necessarily) formed by a small number of neurons, which may be an issue both from the modeling perspective and from the time-series analysis perspective. The intermittent synchronized dynamics studied here naturally occurs in the network of oscillating units because the coupling strength is not very high. However, there may be other factors, which may contribute to the temporal variability of synchrony, such as the noise of different nature and synaptic plasticity (the latter is known to potentially affect temporal synchrony patterns, Zirkle and Rubchinsky, 2020). Nevertheless, given that the network expresses PING gamma, the results of the study are likely to be applicable to the gamma synchronization due to the pyramidal-interneuron gamma mechanism captured by this simple network.

### 4.4. Conclusion

We showed that synaptic changes may alter the temporal patterning of synchronization (and may do so independently of the synchronization strength) in the neural network exhibiting PING gamma rhythm. It was conjectured that this temporal patterning is physiologically important and that the dynamics with short desynchronizations may facilitate formation and break-up of transient neural assemblies (Ahn and Rubchinsky, 2013, 2017). Given the importance of gamma synchronization in facilitation of cognition and the short time scales associated with these phenomena, it is quite plausible that short desynchronization dynamics we observed in the PING gamma network is important for formation of transient neural assemblies and for cognitive phenomena. Stronger local connections and weaker cross-circuit connections between inhibitory and excitatory neurons as well as weaker local and stronger cross-circuit connections between inhibitory and inhibitory neurons (which we found to promote short desynchronizations) may thus play facilitatory role for these phenomena.

**Acknowledgements**

This work was supported by NSF DMS 1813819.

**Data availability statement**

Data sharing is not applicable to this article as no new data were created or analyzed in this study.

Pittman-Polletta, B. R., Kocsis, B., Vijayan, S., Whittington, M. A., and Kopell, N. J. (2015). Brain rhythms connect impaired inhibition to altered cognition in schizophrenia. *Biol. Psychiatry* 77: 1020–1030.

Ratnadurai-Giridharan, S., Zauber, S. E., Worth, R. M., Witt, T., Ahn, S., and Rubchinsky, L. L. (2016). Temporal patterning of neural synchrony in the basal ganglia in Parkinson's disease. *Clin Neurophysiol* 127: 1743–1745.

Rubchinsky, L. L., Park, C., and Worth, R. M. (2012). Intermittent neural synchronization in Parkinson's disease. *Nonlinear Dyn* 68: 329–346.

Salkoff, D., Zagha, E., Yüzgeç, Ö., and McCormick, D. (2015). Synaptic Mechanisms of Tight Spike Synchrony at Gamma Frequency in Cerebral Cortex. *J Neurosci* 35: 10236-10251.

Spellman, T. J., and Gordon, J. A. (2015). Synchrony in schizophrenia: a window into circuit-level pathophysiology. *Curr. Opin. Neurobiol*. 30: 17–23.

Traub, R. D., and Miles, R. (1991). *Neuronal Networks of the Hippocampus*, Cambridge University Press, Cambridge, UK.

Uhlhaas, P. J., and Singer, W. (2010). Abnormal neural oscillations and synchrony in schizophrenia. *Nat. Rev. Neurosci.* 11: 100–113.

Vierling-Claassen, D., Siekmeier, P., Stufflebeam, S., and Kopell, N. Modeling GABA alterations in schizophrenia: a link between impaired inhibition and altered gamma and beta range auditory entrainment. *J Neurophysiol*, 99: 2656-2671.

Wang, X. J., and Buzsáki, G. (1996). Gamma oscillation by synaptic inhibition in a hippocampal interneuronal network model. *J. Neurosci*. 16: 6402–6413.

Woodman, M. M., and Canavier, C. C. (2011). Effects of conduction delays on the existence and stability of one to one phase locking between two pulse-coupled oscillators. *Journal of computational neuroscience*, *31*(2), 401–418. https://doi.org/10.1007/s10827-011-0315-2

Zirkle, J, and Rubchinsky, L. L. (2020) Spike-timing dependent plasticity effect on the temporal patterning of neural synchronization. *Front Comput Neurosci* 14: 52.
16

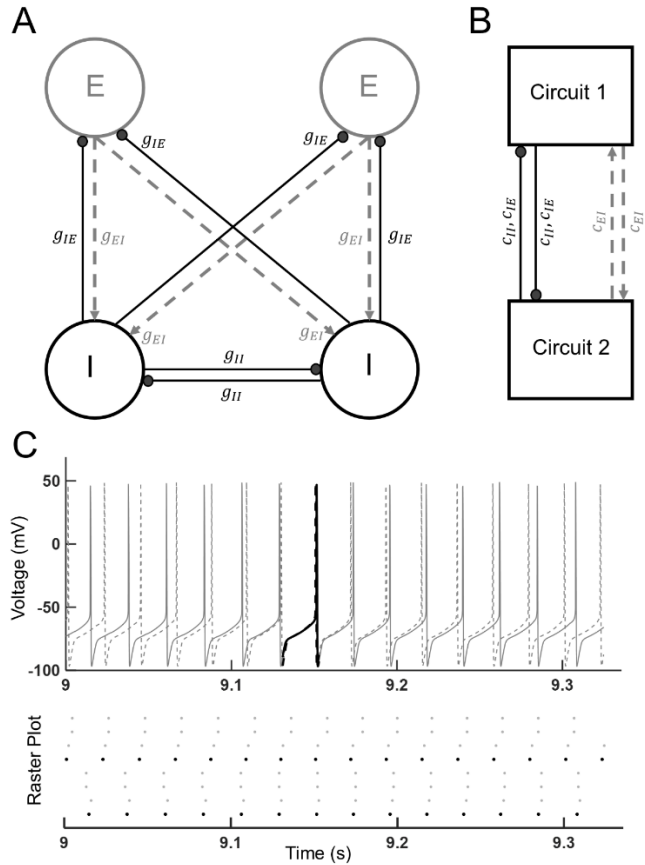

**Figure 1.** Schematic of the model circuitry. A: an individual circuit consists of two excitatory and two inhibitory neurons with excitatory connections $g_{EI}$ (dashed gray with arrow at the end) and inhibitory connections $g_{II}$ and $g_{IE}$ (solid black with circle at the end). B: Full model circuitry has two individual circuits. Two circuits are connected by inhibitory synapses $c_{IE}$ and $c_{II}$ (solid black) and excitatory synapses $c_{EI}$ (dotted gray). There are no mutual connections between excitatory neurons. C: voltage traces of an excitatory neuron from two circuits (grey and black lines) and the raster plot of all the neurons in both networks. The two excitatory neurons with voltage traces above are in black, and the rest of the neurons are in gray.

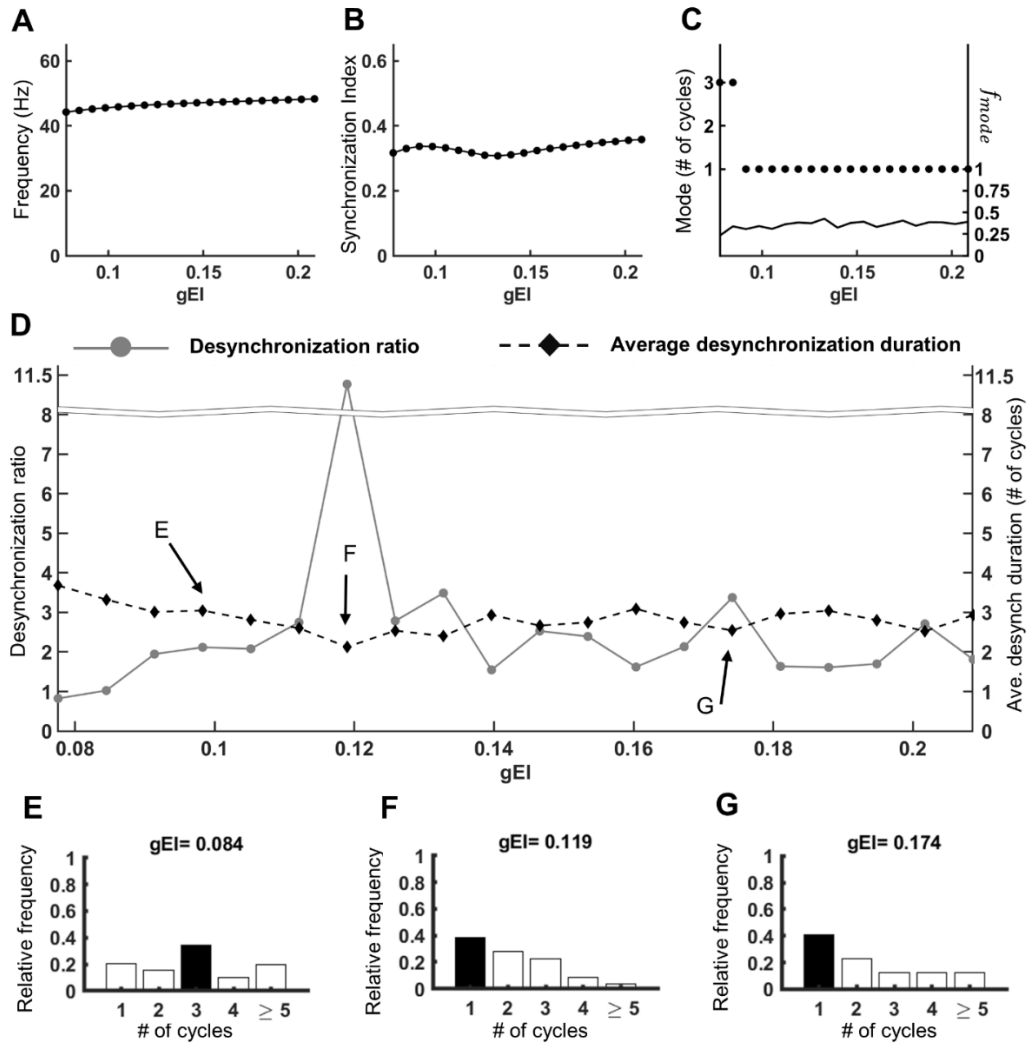

**Figure 2.** Effect of within-circuit excitatory to inhibitory neurons synaptic connection $g_{EI}$ on the temporal patterning of synchronized dynamics. A: Average firing frequency in Hz. B: Synchronization index. C: The mode (# of cycles) of desynchronization durations distribution (black dots) and the frequency of the mode $f_{mode}$ (black curve). D: Average desynchronization duration (# of cycles) in dashed black line with diamonds and desynchronization ratio in solid gray line with circles. Examples of distribution of desynchronization durations are shown in panels E, F, and G; horizontal axis is the duration of desynchronizations as measured in the cycles of oscillations. The mode of desynchronization distribution is highlighted as a black bar in each histogram. The arrows in panel D indicate the cases that correspond to the histograms shown in panels E, F, and G.

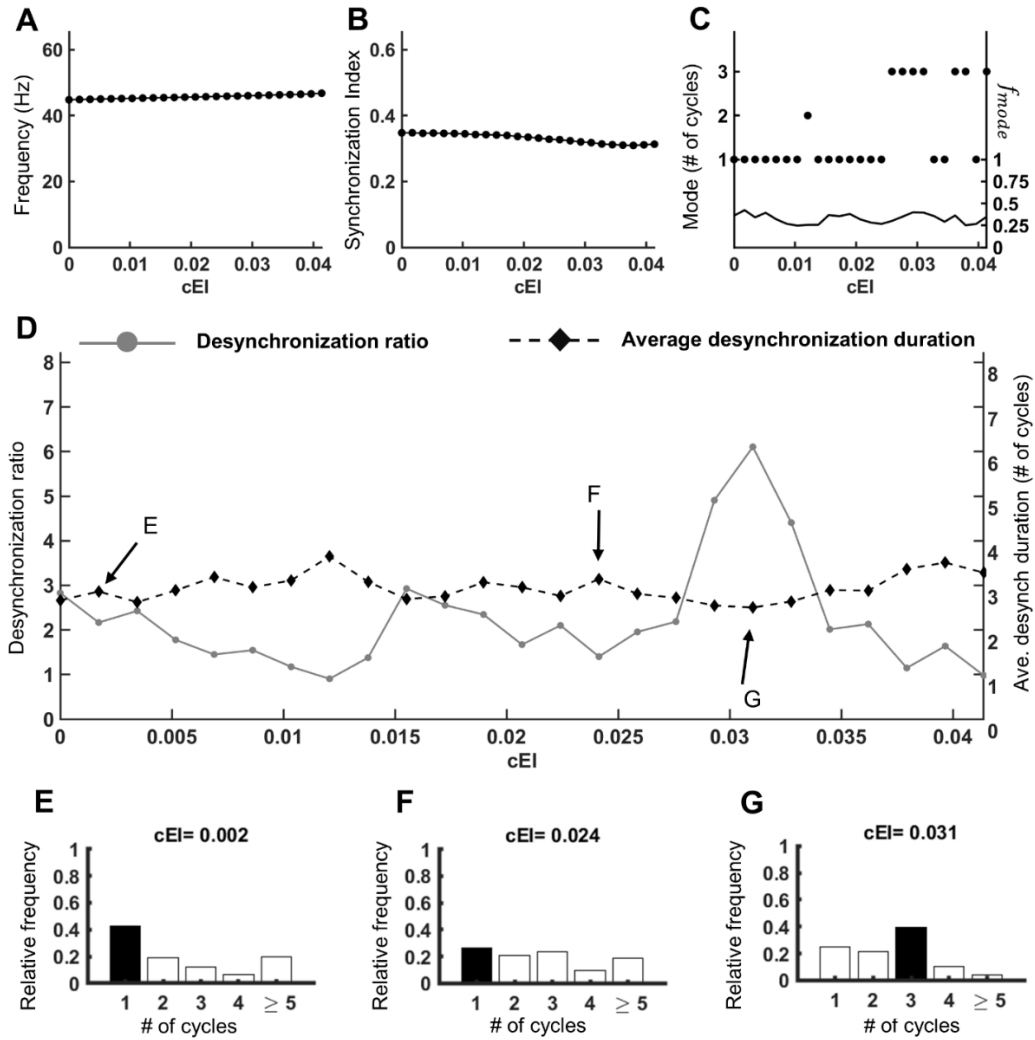

**Figure 3.** Effect of between-circuit excitatory to inhibitory neurons synaptic connection $c_{EI}$ on the temporal patterning of synchronized dynamics. A: Average firing frequency in Hz. B: Synchronization index. C: The mode (# of cycles) of desynchronization durations distribution (black dots) and the frequency of the mode $f_{mode}$ (black curve). D: Average desynchronization duration (# of cycles) in dashed black line with diamonds and desynchronization ratio in solid gray line with circles. Examples of distribution of desynchronization durations are shown in panels E, F, and G; horizontal axis is the duration of desynchronizations as measured in the cycles of oscillations. The mode of desynchronization distribution is highlighted as a black bar in each histogram. The arrows in panel D indicate the cases that correspond to the histograms shown in panels E, F, and G.

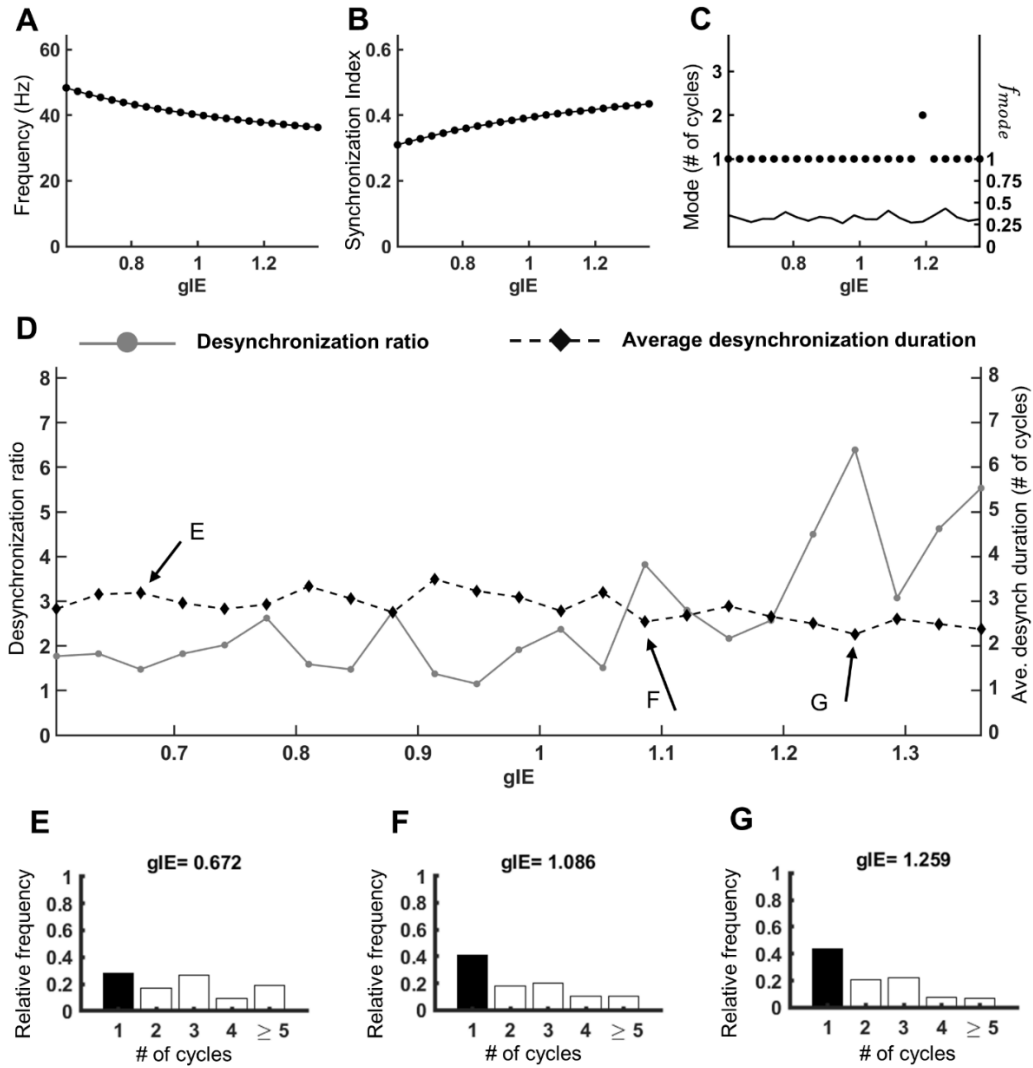

**Figure 4.** Effect of within-circuit inhibitory to excitatory neurons synaptic connection $g_{IE}$ on the temporal patterning of synchronized dynamics. A: Average firing frequency in Hz. B: Synchronization index. C: The mode (# of cycles) of desynchronization durations distribution (black dots) and the frequency of the mode $f_{mode}$ (black curve). D: Average desynchronization duration (# of cycles) in dashed black line with diamonds and desynchronization ratio in solid gray line with circles. Examples of distribution of desynchronization durations are shown in panels E, F, and G; horizontal axis is the duration of desynchronizations as measured in the cycles of oscillations. The mode of desynchronization distribution is highlighted as a black bar in each histogram. The arrows in panel D indicate the cases that correspond to the histograms shown in panels E, F, and G.

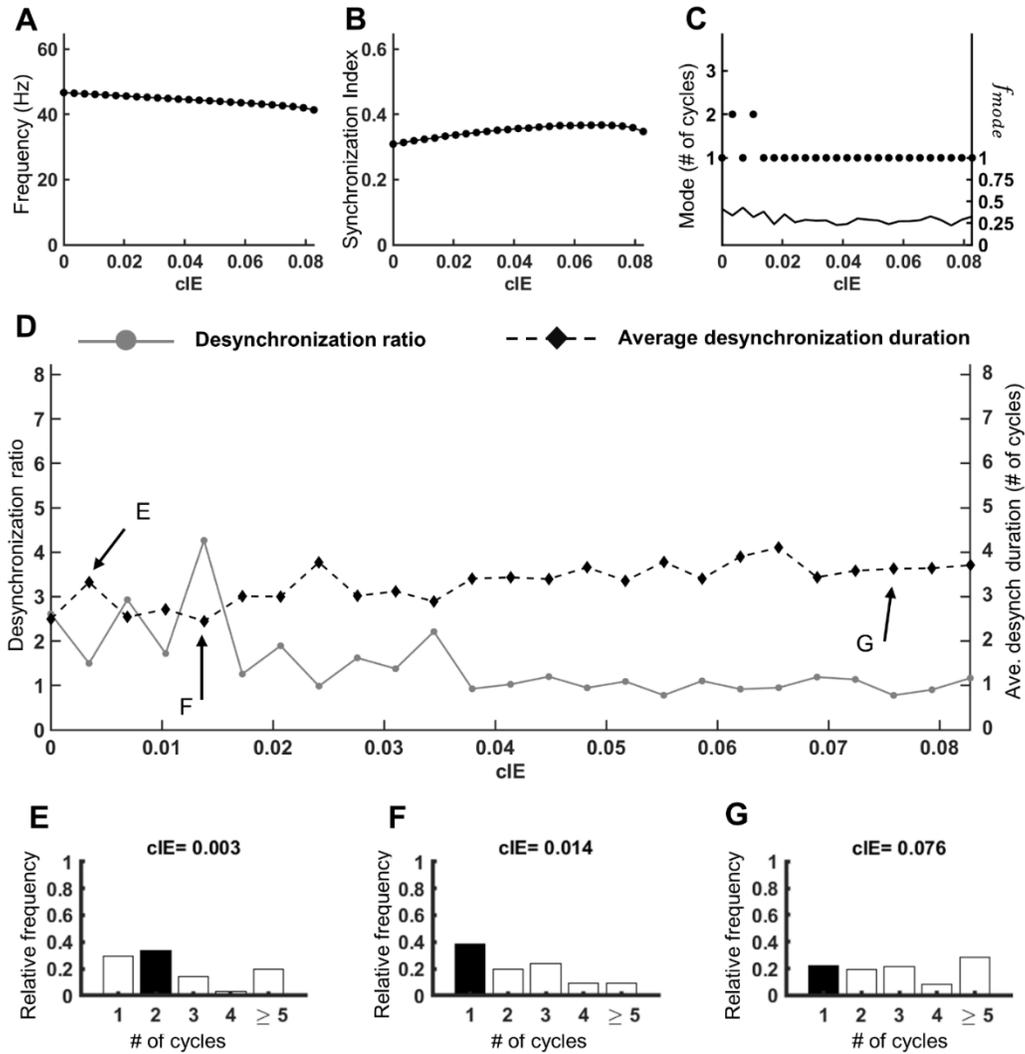

**Figure 5.** Effect of between-circuit inhibitory to excitatory neurons synaptic connection $c_{IE}$ on the temporal patterning of synchronized dynamics. A: Average firing frequency in Hz. B: Synchronization index. C: The mode (# of cycles) of desynchronization durations distribution (black dots) and the frequency of the mode $f_{mode}$ (black curve). D: Average desynchronization duration (# of cycles) in dashed black line with diamonds and desynchronization ratio in solid gray line with circles. Examples of distribution of desynchronization durations are shown in panels E, F, and G; horizontal axis is the duration of desynchronizations as measured in the cycles of oscillations. The mode of desynchronization distribution is highlighted as a black bar in each histogram. The arrows in panel D indicate the cases that correspond to the histograms shown in panels E, F, and G.

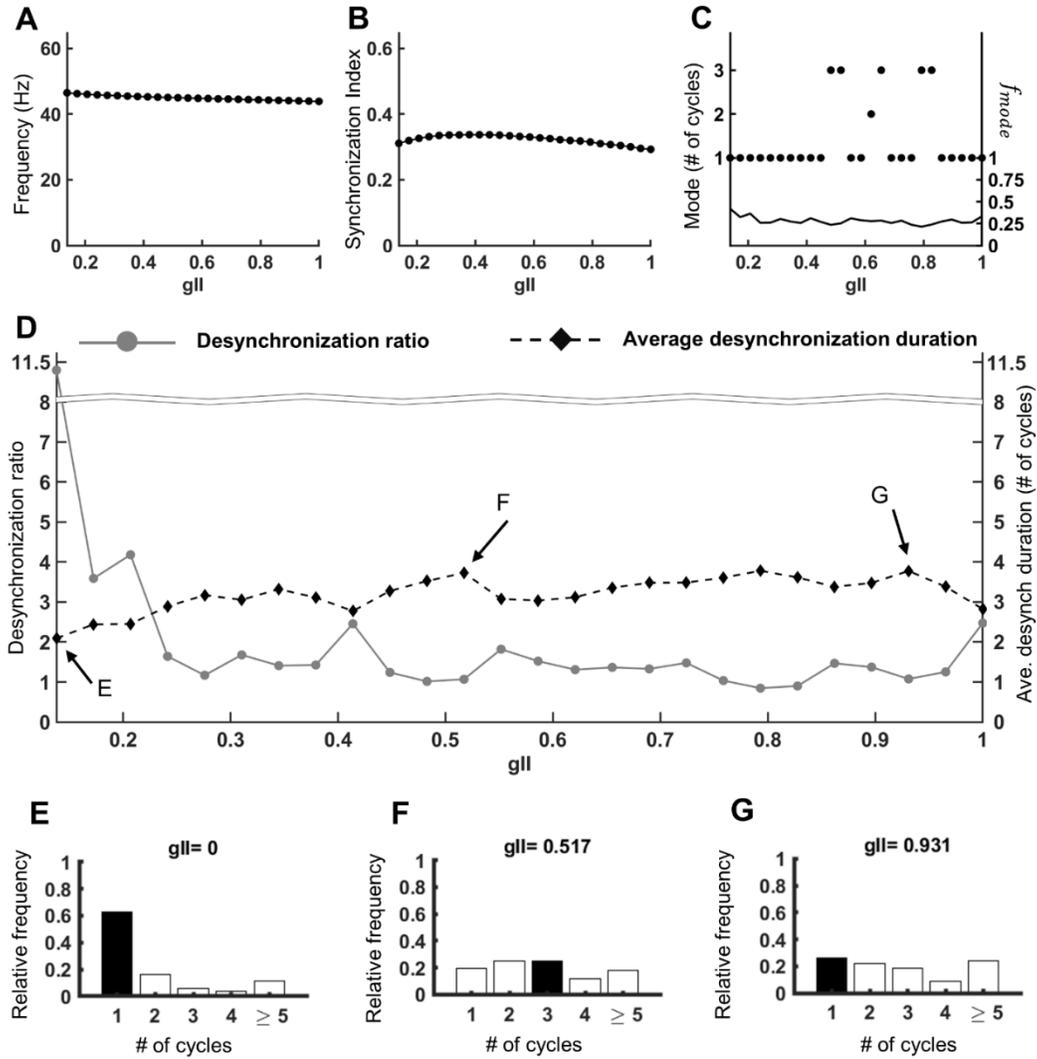

**Figure 6.** Effect of within-circuit inhibitory to inhibitory neurons synaptic connection $g_{II}$ on the temporal patterning of synchronized dynamics. A: Average firing frequency in Hz. B: Synchronization index. C: The mode (# of cycles) of desynchronization durations distribution (black dots) and the frequency of the mode $f_{mode}$ (black curve). D: Average desynchronization duration (# of cycles) in dashed black line with diamonds and desynchronization ratio in solid gray line with circles. Examples of distribution of desynchronization durations are shown in panels E, F, and G; horizontal axis is the duration of desynchronizations as measured in the cycles of oscillations. The mode of desynchronization distribution is highlighted as a black bar in each histogram. The arrows in panel D indicate the cases that correspond to the histograms shown in panels E, F, and G.

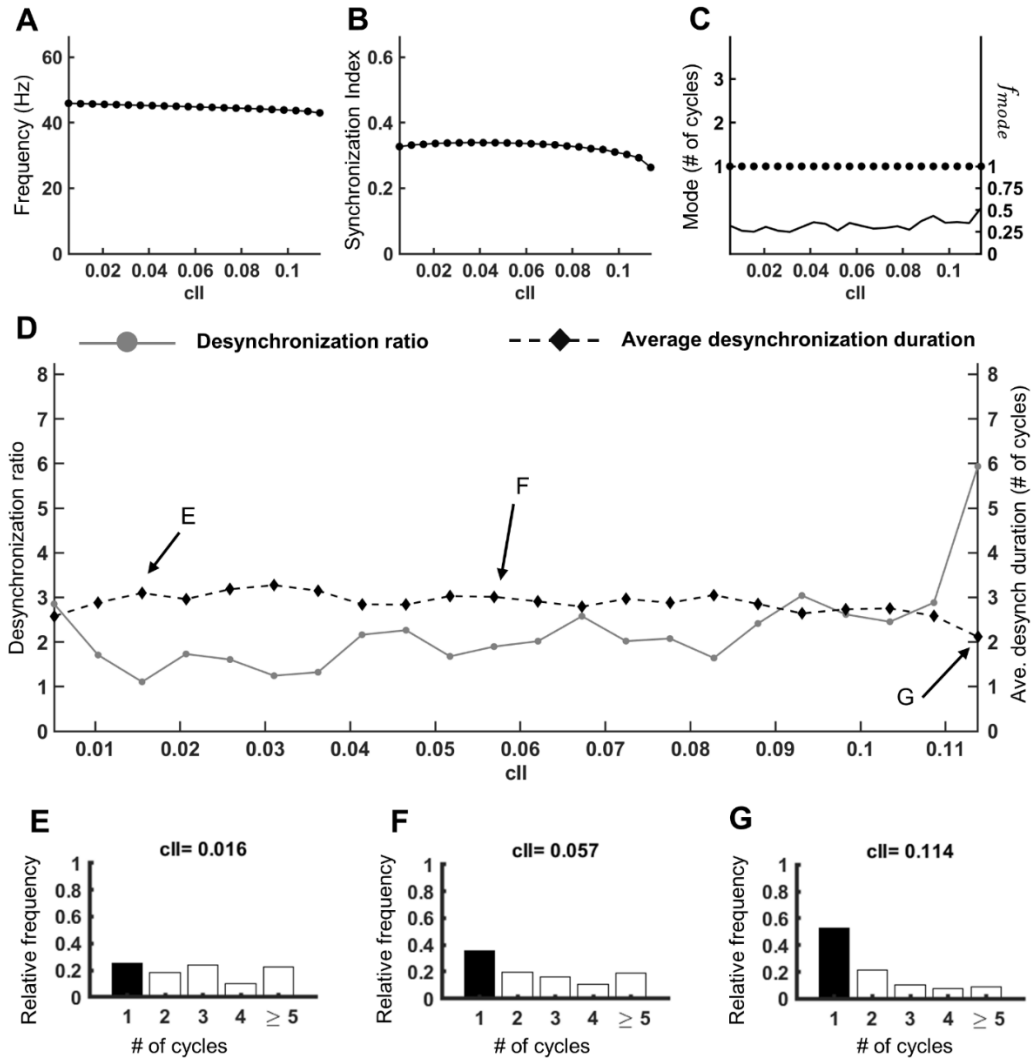

**Figure 7.** Effect of between-circuit inhibitory to inhibitory neurons synaptic connection $c_{II}$ on the temporal patterning of synchronized dynamics. A: Average firing frequency in Hz. B: Synchronization index. C: The mode (# of cycles) of desynchronization durations distribution (black dots) and the frequency of the mode $f_{mode}$ (black curve). D: Average desynchronization duration (# of cycles) in dashed black line with diamonds and desynchronization ratio in solid gray line with circles. Examples of distribution of desynchronization durations are shown in panels E, F, and G; horizontal axis is the duration of desynchronizations as measured in the cycles of oscillations. The mode of desynchronization distribution is highlighted as a black bar in each histogram. The arrows in panel D indicate the cases that correspond to the histograms shown in panels E, F, and G.

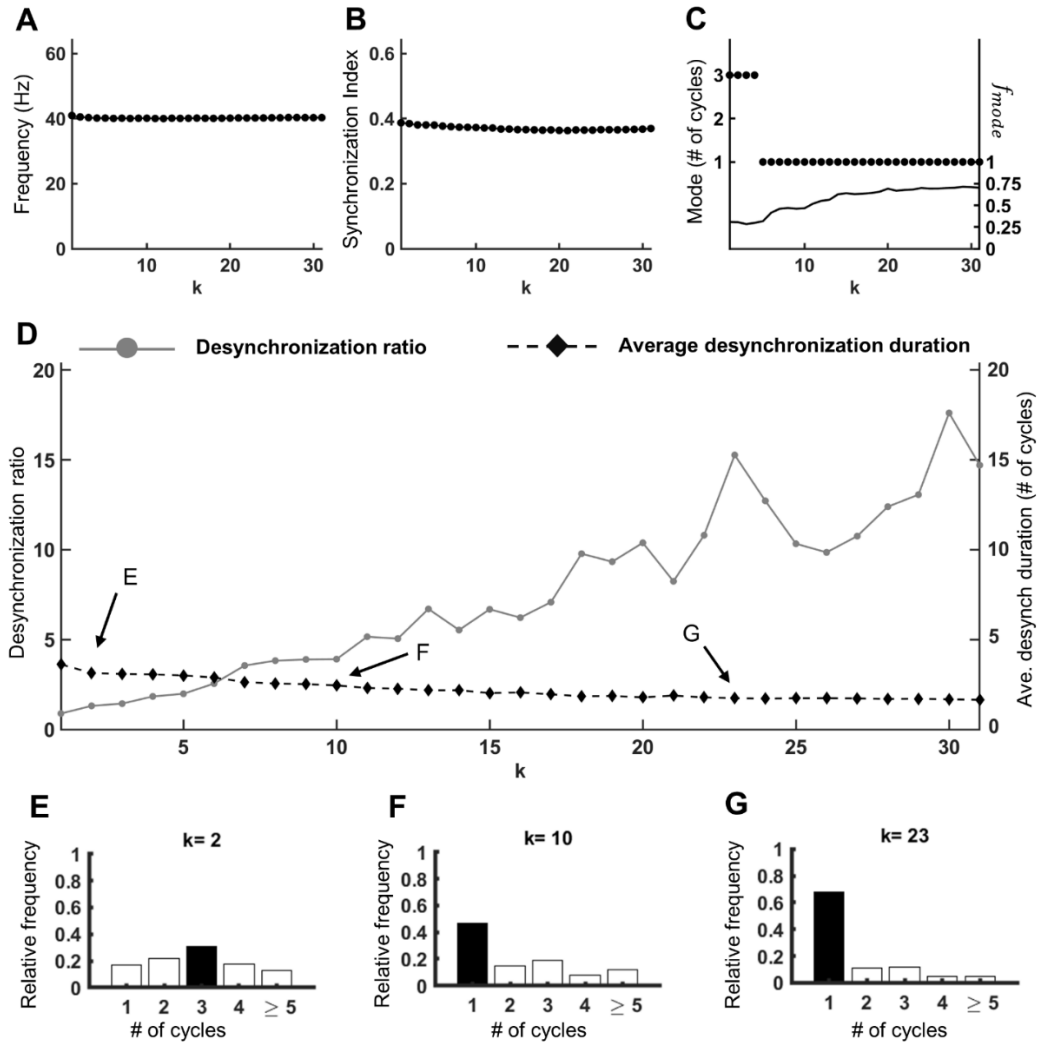

**Figure 8.** Changes in the desynchronization durations can be independent of the frequency and of the average synchronization strength. Local connections between excitatory and inhibitory units $g_{IE}$ and $g_{EI}$ are increased while cross-circuit connections between excitatory and inhibitory cells $c_{IE}$ and $c_{EI}$ are decreased as linear functions of $k$, see Equations (12) and (13) . A: Average firing frequency in Hz. B: Synchronization index. C: The mode (# of cycles) of desynchronization durations distribution (black dots) and the frequency of the mode $f_{mode}$ (black curve). D: Average desynchronization duration (# of cycles) in dashed black line with diamonds and desynchronization ratio in solid gray line with circles. Examples of distribution of desynchronization durations are shown in panels E, F, and G; horizontal axis is the duration of desynchronizations as measured in the cycles of oscillations. The mode of desynchronization distribution is highlighted as a black bar in each histogram. The arrows in panel D indicate the cases that correspond to the histograms shown in panels E, F, and G.

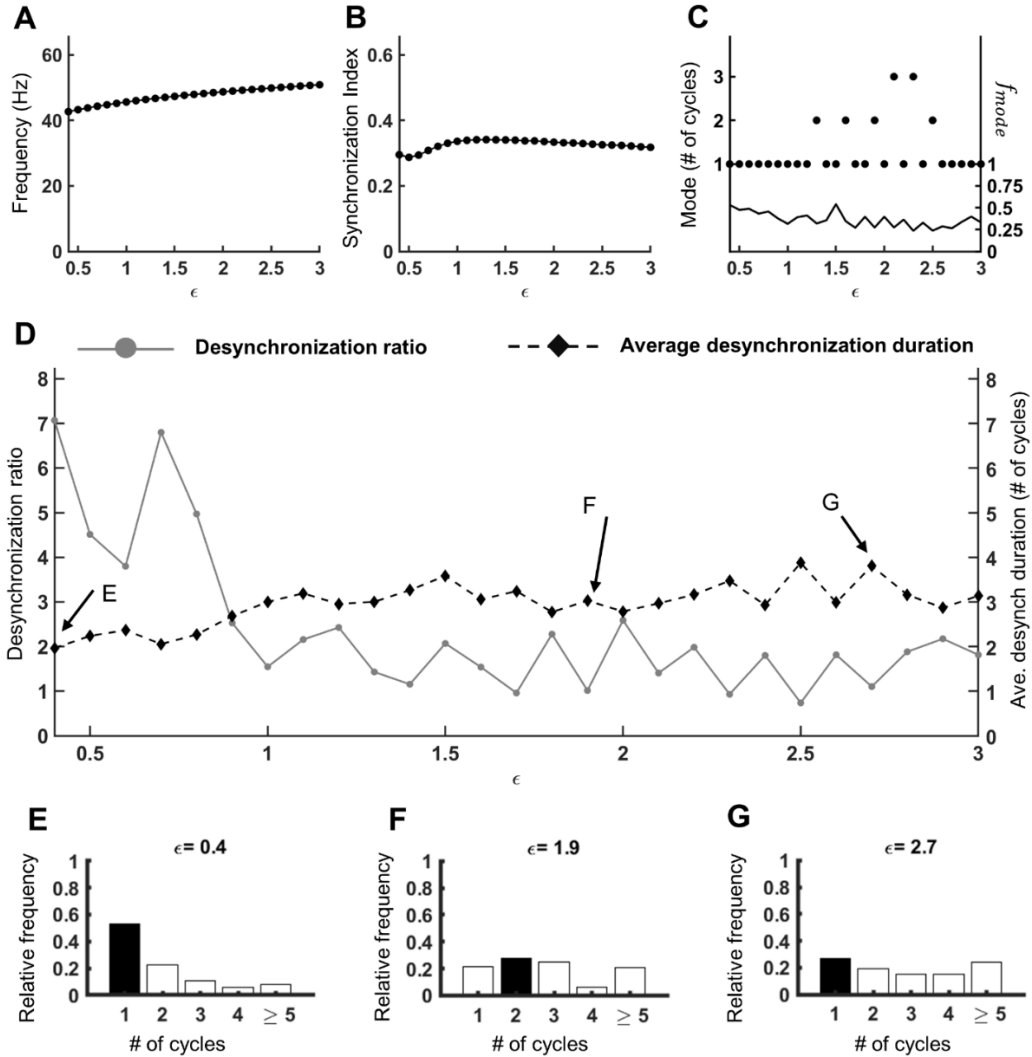

**Figure 9.** Effect of the peak value of activation time constant of potassium channel on the temporal patterning of synchronized dynamics. The peak value of activation time constant is written as function of $\varepsilon$, see Equations (14) and (15)). A: Average firing frequency in Hz. B: Synchronization index. C: The mode (# of cycles) of desynchronization durations distribution (black dots) and the frequency of the mode $f_{mode}$ (black curve). D: Average desynchronization duration (# of cycles) in dashed black line with diamonds and desynchronization ratio in solid gray line with circles. Examples of distribution of desynchronization durations are shown in panels E, F, and G; horizontal axis is the duration of desynchronizations as measured in the cycles of oscillations. The mode of desynchronization distribution is highlighted as a black bar in each histogram. The arrows in panel D indicate the cases that correspond to the histograms shown in panels E, F, and G.

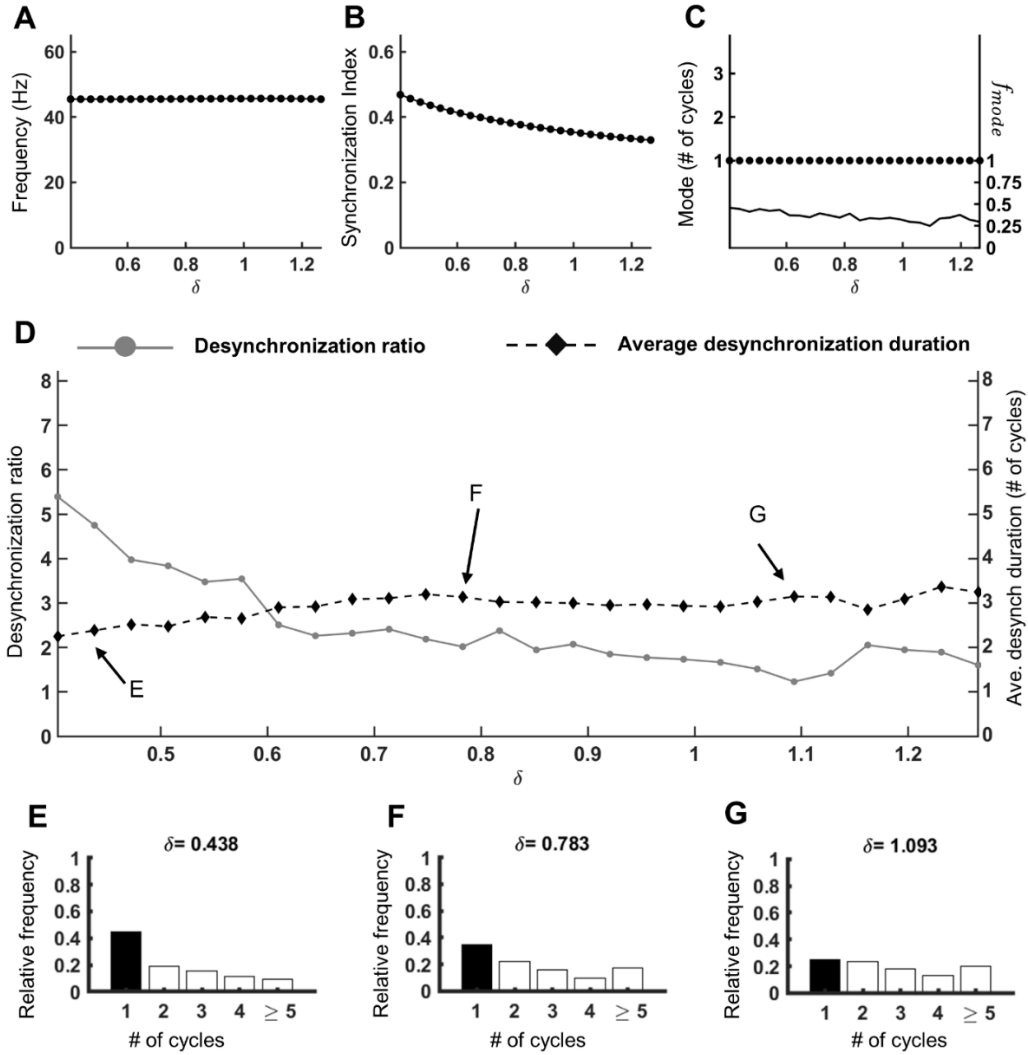

**Figure 10.** Effect of the width of voltage-dependence of the activation time constant of potassium channel on the temporal patterning of synchronized dynamics. The width of voltage-dependence of the activation time constant is parametrized by δ, see Equations (14) and (15). A: Average firing frequency in Hz. B: Synchronization index. C: The mode (# of cycles) of desynchronization durations distribution (black dots) and the frequency of the mode $f_{mode}$ (black curve). D: Average desynchronization duration (# of cycles) in dashed black line with diamonds and desynchronization ratio in solid gray line with circles. Examples of distribution of desynchronization durations are shown in panels E, F, and G; horizontal axis is the duration of desynchronizations as measured in the cycles of oscillations. The mode of desynchronization distribution is highlighted as a black bar in each histogram. The arrows in panel D indicate the cases that correspond to the histograms shown in panels E, F, and G.